\documentclass[aip,apl,amsmath,amssymb,reprint,superscriptaddress]{revtex4-1}
\UseRawInputEncoding
\usepackage{graphicx}
\usepackage{amsmath}
\usepackage{dcolumn}
\usepackage{bm}
\usepackage{bbold}
\usepackage{wrapfig}
\usepackage{epsfig}
\usepackage{sidecap}
\usepackage{nomencl}

\bibstyle{apsrev4-1}

\begin{document}

\title{Ferroelectric phase transition and crystal asymmetry monitoring of $SrTiO_3$ using quasi $TE_{m,1,1}$ and quasi $TM_{m,1,1}$ modes}

\author{M. A. Hosain}
\affiliation{ARC Centre of Excellence for Engineered Quantum Systems, School of Physics, University of Western Australia, 35 Stirling Highway, Crawley WA 6009, Australia.}

\author{J-M. Le Floch}
\affiliation{MOE Key Laboratory of Fundamental Physical Quantities Measurement, School of Physics, Huazhong University of Science and Technology, Wuhan 430074, Hubei, China.}
\affiliation{ARC Centre of Excellence for Engineered Quantum Systems, School of Physics, University of Western Australia, 35 Stirling Highway, Crawley WA 6009, Australia.}

\author{J. F. Bourhill}
\affiliation{ARC Centre of Excellence for Engineered Quantum Systems, School of Physics, University of Western Australia, 35 Stirling Highway, Crawley WA 6009, Australia.}

\author{J. Krupka}
\affiliation{Department of Electronics and Information Technology, Institute of Microelectronics and Optoelectronics, Warsaw University of Technology, Koszykowa 75, 00-662 Warszawa, Poland.}

\author{M. E. Tobar}
\email{michael.tobar@uwa.edu.au}
\affiliation{ARC Centre of Excellence for Engineered Quantum Systems, School of Physics, University of Western Australia, 35 Stirling Highway, Crawley WA 6009, Australia.}


\begin{abstract}
Dielectric spectroscopy of $SrTiO_3$ single crystal over a broad range of microwave frequency using quasi $TE_{m,1,1}$ and quasi $TM_{m,1,1}$ modes reveals crystal asymmetry from typical measurement of $Q$-factor, transmission or frequency characteristics in continuous cooling down to a few Kelvin. The properties of the modes due to the crystal asymmetry is validated by implementing a quasiharmonic phonon approximation. The observed ferroelectric phase transition temperature is around $51~K$, and quantum-mechanical stabilization of the paraelectric phase arises below $5~K$ with very high permittivity. Also, an antiferroelectric distortive transition was indicated at $105~K$. Landau's theory of correlation length supports the observation of an extra loss term so the transition may be identified near the $Q$-factor maxima or transmission maxima depending on the other loss terms present in the cavity. Thus, the ferroelectric phase transition with respect to temperature may be identified when this extra-loss term causes a discontinuity in the derivative of the temperature characteristic near the minimum of total cavity loss (maxim Q-factor or maximum transmission temperature characteristic). This temperature is confirmed by transmission amplitude variation under 200 V dc electric field showing existence of the soft-mode. These measurements support a typical polarization model and explicit temperature dependency of the soft-mode incorporating an imaginary frequency.
\end{abstract} 
\maketitle

\subsection{Introduction:}
As a microscopic view, crystal structure and collective dynamics of lattice (phonon vibration) have an important role on the dielectric polarization and the possibility of ferroelectric (FE)\nomenclature{FE}{Ferroelectric} phase transitions\cite{Polariz,DisplaciveFE}. In this work we implemented a probing process using the quasi $TE_{m,1,1}$ (transverse electric) and quasi $TM_{m,1,1}$ (transverse magnetic) dielectric resonance microwave modes\cite{WG,LeFloch,Krupka,Krupka1,Konopka} to monitor dielectric resonator crystal structure asymmetry and such a FE phase transition with respect to temperature. The excited dielectric resonance modes are calculated as a solution of Bessel function with indices $(m,n,l)$ of $TE_{m,n,l}$ and $TM_{m,n,l}$ modes as m for azimuthal variation, n for radial variation and l for axial variation. The fact that the phonon quasi-harmonic approximation  allows the analysis in a simple perturbation theory. Any change in the crystal lattice changes the line width of the modes, hence, is being monitored in the continuous cooling. We measured the phase transition temperature and structural changes of a white opaque $SrTiO_3~(STO)$ crystal\cite{Anna} specimen using this process. The asymmetry in the crystal unit due to soft mode phonons, which create a dielectric anomaly in macroscopic state contributing to imaginary part of relative permittivity\cite{DisplaciveFE} and immediately changes the observable quantities like Q-factor, transmission or frequency shift with respect to temperature. Hence the excited dielectric resonance mode in the microwave regime acts as a frequency transducer of a conversion of the change of the lattice structure of the crystal into the most precise physical and measurable quantity which is frequency, thus works as a powerful sensitive tool for monitoring the nature of structural asymmetry of crystal unit and phase transitions.\

The substance $SrTiO_3$ crystallizes in a perovskite structure, according to the space group Pm3m with lattice constant $3.905$~\r{A} in cubic symmetry, with a $Ti^{4+}-O^{2-}$ distance of $1.952$~\r{A}\cite{STO-Structure,STO-CrystField1}. FE states\cite{DisplaciveFE} arise from a para-electric (PE) phase while cooling as a displacive structural phase change of the crystal\cite{OriginFE}. The FE free energy equilibrium state's qualitative agreement with dielectric polarization is explained by Landau theory\cite{Landau}. This phenomenological parameter\cite{Landau}, the dielectric polarization\cite{Polariz} reveals spontaneous polarization as an order-disorder parameter of double-well potential\cite{DisplaciveFE,DisplaciveFE_Book} as the signature of FE phase initiating a rapid decrease of the measured resonant mode Q-factor.\
 
Different types of experiments have been conducted on this well known single crystal $SrTiO_3$. Pressure-induced antiferrodistortive cubic-to-tetragonal phase transition in $SrTiO_3$ was observed by Weng et al. at ambient temperature\cite{PressAntiFerro}. Salje et al. demonstrated that upon decreasing temperature instead of pressure, STO at ambient pressure undergoes a transition at $105~K$  characterized by the same type of symmetry breaking\cite{PressAntiFerro1}. Further reduction of the temperature results in another transition, which has no counterpart for the same system under pressure at ambient temperature. Furthermore, K. A. M$\ddot{u}$ller and H. Burkard have shown that a quantum-mechanical stabilization of the paraelectric phase arises in $SrTiO_3$ below 4 K, with a very high dielectric constant as an intrinsic feature\cite{MullerQuanPE}. In a temperature dependent polarization experiment studies on strained STO films displays a ferroelectric polarization and hysteresis loop at room temperature\cite{RoomTemFE}. By a direct comparison of the strained and strain-free SrTiO3 films using dielectric, ferroelectric, Raman, nonlinear optical and nanoscale piezoelectric property measurements, Jang et al. concluded that all $SrTiO_3$ films and bulk crystals are relaxor ferroelectrics, and the role of strain is to stabilize longer-range correlation of preexisting nanopolar regions\cite{ComparisonFE}.

The crystal structure changes in FE phase transitions like cubic to tetragonal, and the center-symmetric ion in the octahedral structure of the crystal unit is substantially displacive instead of diffusive\cite{STO-BTO-lattice,STO-FewFE}. The quasi $TE_{m,1,1}$ modes and quasi $TM_{m,1,1}$ modes are very sensitive to this FE phase transition produced spontaneous polarization\cite{SpontPola} due to the soft-mode producing a displacement of the center-symmetric ion of the crystal unit\cite{STO-Struc}. The spontaneous polarization\cite{SpontPola,OriginFE,STO-FewFE} contributes to the net dielectric polarization of the crystal showing a large shift of quari $TE$ and $TM$ mode resonant frequencies, apparent in the frequency-temperature characteristic curves, and the cause of rapid decrease of Q-factor\cite{LossFE}. The observed Q-factor variation is the change of line width of the modes due to damping of spontaneous polarization created dipole and used can be used to monitor the phase transition temperature of the crystal in the cooling down to few Kelvin.\
  
Quasi-harmonic approximation of phonon\cite{DisplaciveFE,PseudoFE} is required to adapt with the anharmonic states of FE perturbation. Soft-mode\cite{DisplaciveFE_Book,SoftMode} and Hard-mode\cite{DisplaciveFE_Book,Hardmode} concepts of mode frequency shifts are described to explain the measurement of phase transition temperatures of resonator, and significantly monitored in terms of observable quantity Q-factor, transmission amplitude, and frequency shift in our experiments of dielectric spectroscopy.

\subsubsection{Phase Transition Mechanism with Temperature:}

\begin{figure}[h!]
\centering
\includegraphics[width=0.5\textwidth]{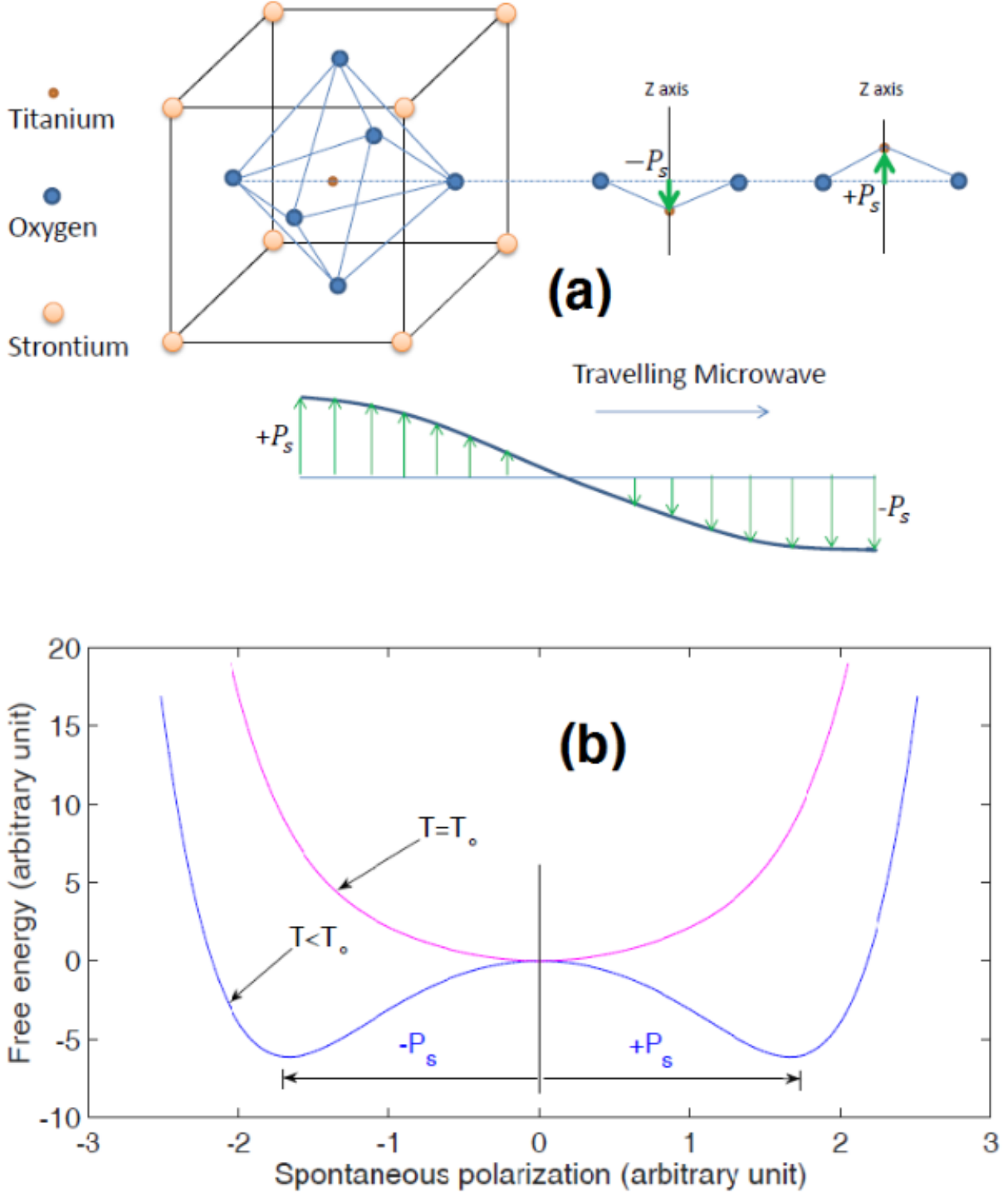}
\caption[Gibbs free energy with phase transition]{\label{STO} Characteristics of Gibbs free energy $\phi$ in phase transitions showing spontaneous polarization $P_s$ as a solution (using Landau theory) of one dimensional dielectric polarization $P_x$ equation\cite{Landau,DisplaciveFE} (with crystal symmetry constants $\alpha_\circ, \alpha_{11}$ and $\alpha_{111}$)\cite{DisplaciveFE}:
 $\phi= \phi_\circ + \frac{\alpha_\circ (T-T_\circ)P^2_x}{2}+\frac{\alpha_{11} P^4_x}{4}+\frac{\alpha_{111} P^6_x}{6}$
The free-energy term $\phi$ stands as the sum of PE free-energy $\phi_\circ$ and FE energy arises from small displacement of central ion of octahedral structure $TiO_6$ (Fig.\ref{STO}a) due to cooling. There is a continuous phase transition (second-order) at $T=T_\circ$. At the lower temperature ($T\leq T_\circ$), $P_s$ has the temperature dependence (calculated neglecting the higher order terms than $P_x^4$) as:
 $P_s =\pm \bigg\lbrack \frac{\alpha_\circ (T- T_\circ)}{\alpha_{11}}\bigg\rbrack^\frac{1}{2}$. A first order phase transition could appear at $T \geq T_\circ \leq T_c$ with $\alpha_{11} < 0$ showing a discontinuity within this temperature range in some crystal.
}
\end{figure}

 Landau theory describes that the particle correlation length characteristics with respect to phenomenological parameter (polarization)\cite{Landau,DisplaciveFE,Relaxor,Curie,Soft,SDriven} reveals equilibrium state with Gibbs free energy $\phi$ as ${\frac{\partial \phi}{\partial P(T)}}=0$ at the ferroelectric phase transition temperature $T_\circ$ of the crystal\cite{Polariz,Landau,DisplaciveFE,Relaxor,SDriven}. Applying this equilibrium condition of FE free energy, the spontaneous polarization $P_s$ shows two minima at $P_x=\pm P_s$ and a single maxima at $P_s=0$ where $T<T_\circ$ and $T=T_\circ$ respectively. On the other hand, only a single minima appears when $T\geq T_\circ$ (Fig.\ref{STO}b).

In the phenomena of ionic polarization due to displacement (Fig.\ref{STO}a), core-shell interaction \cite{Polariz,STO-CoreShell,Fluctu,STO-QuOxyPola,STO-Watson} among individual ion $Ti^{4+}$ and oxygen in the $TiO_6$ cluster, with interacting coupling constant $g_2$ of attractive coulomb interactions, and coupling constant $g_4$ directly related to oxygen polarizability\cite{STO-CoreShell} are considered for the free-energy of the system $\phi$ as a contribution of rigid ion potential $V_\circ$, polarizability potential $V_w$, and temperature dependent SPA (Self-consistent Phonon Approximation)\nomenclature{SPA}{Selfconsistent Phonon Approximation}. Hence, the free energy adopts the form\cite{Polariz}:
 
 \begin{equation}
\label{eq:Freeenergy3}
 \frac{\partial \phi}{\partial \langle W_1\rangle_T}= \langle W_1\rangle_T \left[ g_2+g_4 \langle W_1 \rangle ^2_T+ 3g_4 \langle W_1^2 \rangle_T\right]
\end{equation}
 which is zero for polarization displacement $\langle W_1 \rangle _T=0$ and indicates paraelectric phase; or, $\langle W_1\rangle_T=\pm \left[-\frac{g_2}{g_4}-3\langle W_1^2\rangle_T\right]^{\frac{1}{2}}=\pm\left[2\left(\langle W_1^2\rangle_{T_\circ} -\langle W_1^2\rangle_T\right)\right]^\frac{1}{2}$ in ferroelectric phase. This is also analogous to Landau theory\cite{Polariz,Landau,DisplaciveFE}.
 
\subsubsection{Quasiharmonic approximation and explicit temperature dependent imaginary frequency:}
  The energy associated with the lattice vibrations is usually expressed in terms of the Hamiltonian of phonon $H_{ph}$. For a crystal containing Z atoms in each unit cell, the $3Z$ phonon branches are labelled by $\nu$. To make the equation compact, the mode label $(\hat{q},\nu)$ is denoted by wave vector $q$ and $({-\hat{q},\nu})$ is denoted by wave vector $-q$. The amplitude of each phonon of wave vector $\tilde{q}$ is represented by the mass weighted normal-coordinate $Q_q$ in the lattice\cite{DisplaciveFE}. To give an effective Hamiltonian from general interaction, the anharmonic interaction term $V_4$ of $4^{th}$ order in normal-coordinate thermal average are considered in the form of the Hamiltonian\cite{DisplaciveFE,DisplaciveFE_Book1}
\begin{eqnarray}
\label{eq:Freeenergy4}
 \lefteqn {H_{ph}=\frac{1}{2}\sum_q \omega_q^2Q_qQ_{-q}}\\\nonumber &+ &\frac{1}{4}\sum_q\sum_{q'} V_4(q,-q,q',-q') \langle Q_{q'}Q_{-q'}\rangle Q_q Q_{-q}   
 \end{eqnarray}
  
 Hence the Hamiltonian can be rearranged considering anharmonic interactions with a set of renormalized phonon frequencies $\tilde{\omega}_q$ as in the following\cite{DisplaciveFE}:
 
\begin{eqnarray}
\label{eq:Freeenergy6}
\lefteqn{H_{ph}=\frac{1}{2}\sum_q \bigg\lbrack  \omega_q^2 +\frac{k_B T}{2}\sum_{q^\prime}} \nonumber \\& & V_4(q,-q,q^\prime,-q^\prime)/{\tilde{\omega}_{q^\prime}^2} \bigg\rbrack\ Q_q Q_{-q} =\frac{1}{2}\sum_q \tilde{\omega}_q^2Q_qQ_{-q}
\end{eqnarray}

Now, \textbf{\textit{the renormalized frequencies have an explicit temperature dependence}} as directly extracted from equation-\ref{eq:Freeenergy6}:

\begin{eqnarray}
\label{eq:Freeenergy7}
 \tilde{\omega}_q^2=\omega_q^2 +\frac{k_BT}{2} \sum_{q^\prime} V_4(q,-q,q^\prime,-q^\prime)/{\tilde{\omega}_{q^\prime}^2}
\end{eqnarray}

This equation has a self-consistent set of solutions for the renormalized frequencies\cite{DisplaciveFE_Book1} and this quasi harmonic approximation application is also called pesudo-harmonic approximation\cite{PseudoFE}.\

According to the soft mode behaviour, the crystal has a small displacive distortion (Fig.\ref{STO}a) corresponding to renormalized frequency  that can be expressed in a normal mode coordinate\cite{DisplaciveFE_Book} and viewed as a small modification of its symmetry structure. This is only possible if $\omega_q^2<0$ (see equation-\ref{eq:Freeenergy7}). Hence, $\tilde{\omega}_q^2=0$ for a zero displacive distortion at a certain temperature $T_\circ$ known as the FE phase transition temperature, and eventually the renormalized phonon frequency $\tilde{\omega}_q$ has an imaginary value at $T<T_\circ$ temperature \cite{DisplaciveFE}. From equation-\ref{eq:Freeenergy7}:

\begin{eqnarray}
\label{eq:Freeenergy8}
 T_\circ =-2\omega_q^2 \bigg/\bigg\lbrack k_B\sum_{q^\prime} V_4(q,-q,q^\prime,-q^\prime)/{\tilde{\omega}_{q^\prime}^2}\bigg\rbrack
\end{eqnarray}

and thus 

\begin{equation}
\label{eq:Freeenergy9}
 \tilde{\omega}_q^2 =\bigg\lbrack\frac{k_B}{2}\sum_{q^\prime} V_4(q,-q,q^\prime,-q^\prime)/{\tilde{\omega}_{q^\prime}^2}\bigg\rbrack(T-T_\circ)
\end{equation}

The dependence of lattice energy at low temperature $(T<T_\circ)$ can be approximated from caption equation of fig.\ref{STO} and also depth of double-well potential $V(P_s)$ (Fig.\ref{STO}b) as described earlier. Now, the coefficient $\alpha_\circ (T-T_\circ)$ is equivalent to the square of the  imaginary harmonic frequency $\tilde{\omega}_q^2$ of temperature dependent mode\cite{DisplaciveFE}. Also this frequency becomes positive on heating $(T>T_\circ)$ from its value $\tilde{\omega}_q^2=0$ at $T=T_\circ$ as $V_4>0$.

\subsection{Measurements and results:} 

 \subsubsection{Dielectric measurements of $SrTiO_3$ (STO) crystal:}
 
\begin{figure}[t!]
\centering
\includegraphics[width=0.5\textwidth]{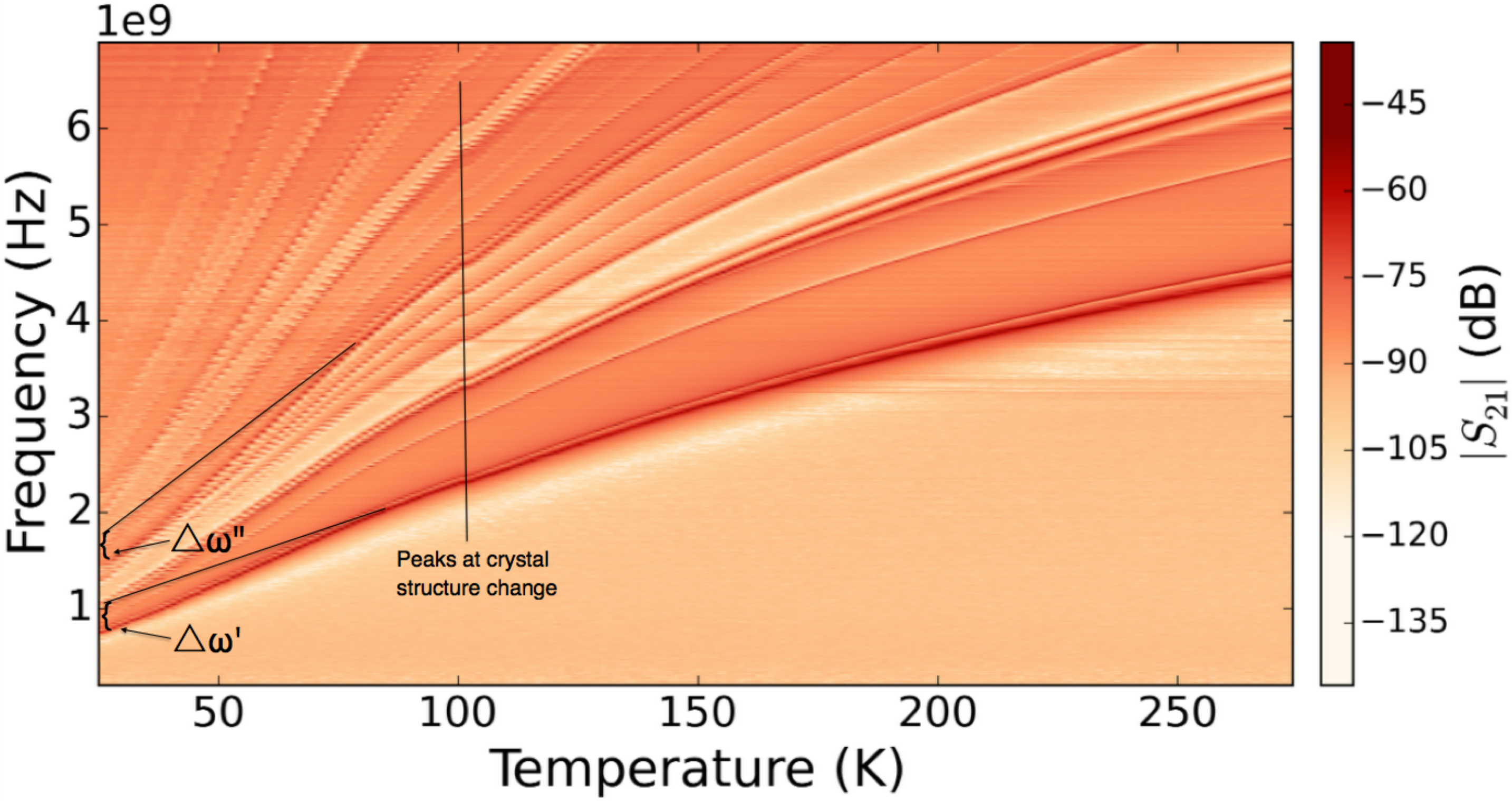}
\caption[Frequency shift characteristics of a bunch of dielectric resonance modes of STO under continuous cooling]{\label{STO1} Frequency shift characteristics of a bunch of dielectric resonance modes of STO under continuous cooling. $\bigtriangleup\omega'$ and $\bigtriangleup\omega''$ are marked for frequency shift due to soft mode of incipient FE phase transition.}
\end{figure}

\begin{figure}[t!]
\centering
\includegraphics[width=0.5\textwidth]{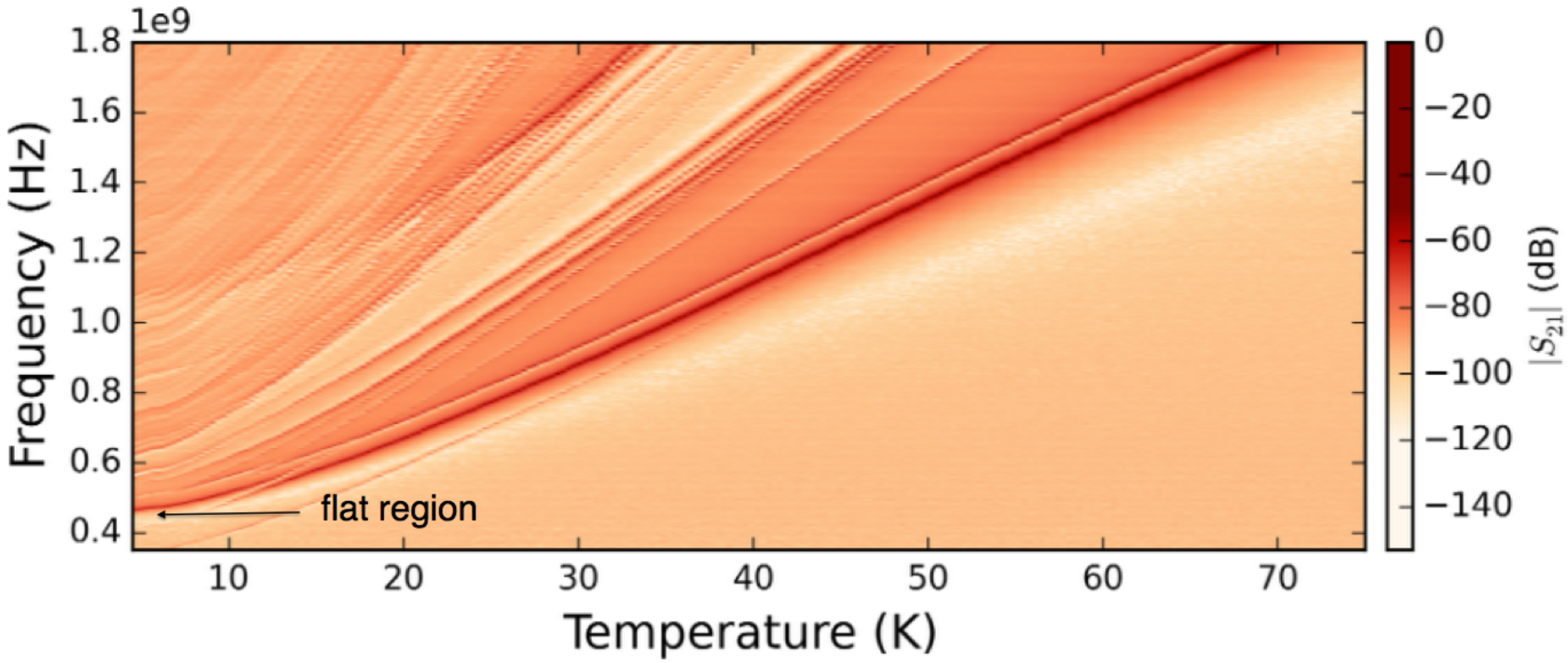}
\caption[Frequency shift characteristics of a bunch of dielectric resonance modes of STO at cryogenic temperature]{\label{STO2} Frequency shift characteristics of a bunch of dielectric resonance modes of STO under continuous cooling at very low temperature. Quantum criticality arises after $4~K$ showing a flat region.}
\end{figure}

\begin{figure}[t!]
\centering
\includegraphics[width=0.5\textwidth]{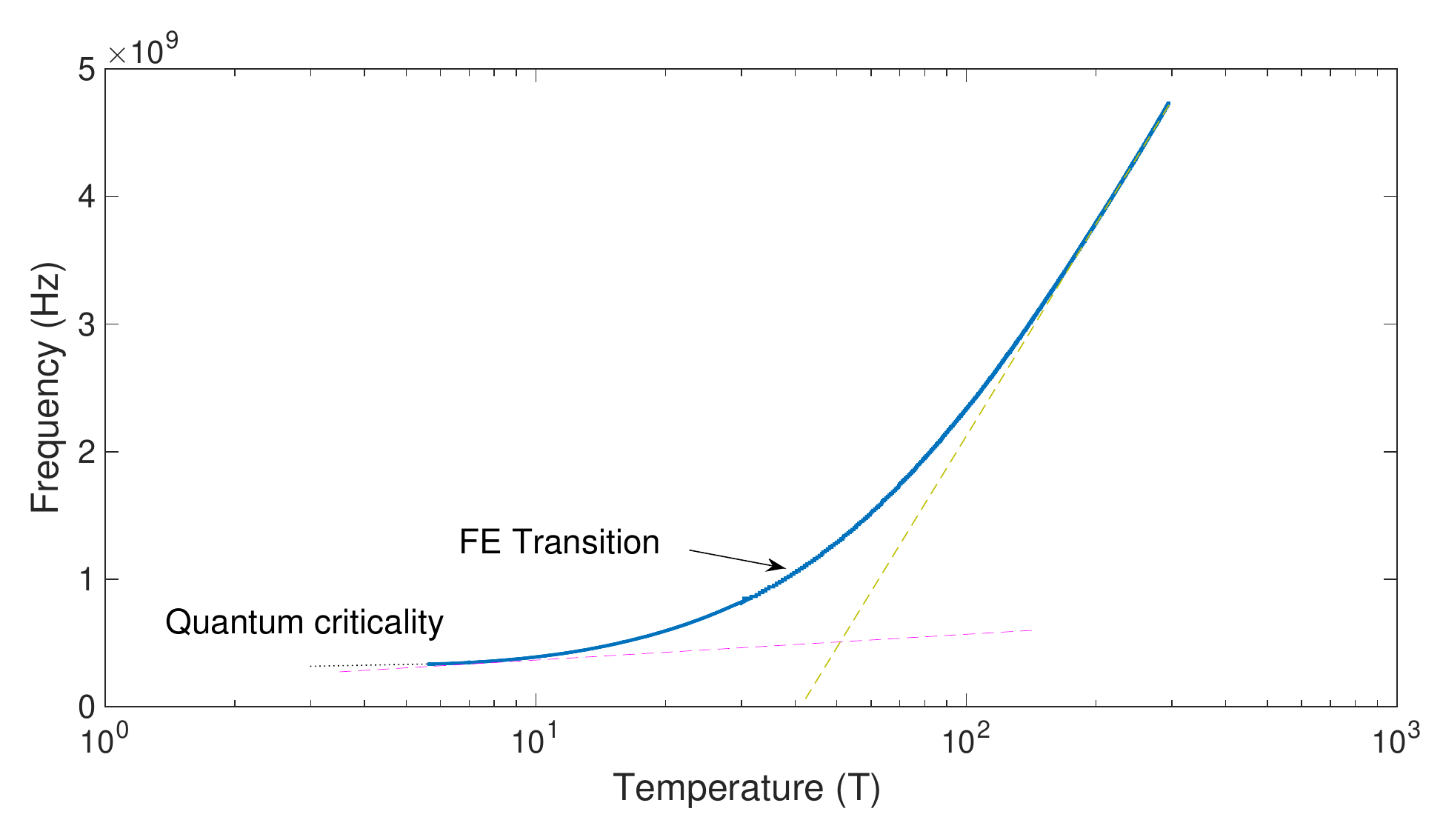}
\caption[Frequency shift characteristic of $SrTiO_3$]{\label{STO5} Frequency shift characteristic of selected mode $TE_{2,1,1}$. The curve turns to a flat region after $4~K$.}
\end{figure}

A cylindrical STO\nomenclature{STO}{Strontium Titanium Oxide} crystal specimen of diameter $3.27~mm$ and height $3.66~mm$ was used to measure dielectric properties. The dielectric resonance mode families, quasi $TE_{m,1,1}$ and quasi $TM_{m,1,1}$ microwave modes  in the range of frequency $1~GHz$ to $11~GHz$ were estimated with azimuthal variation $m=1$ to $4$ by using computer simulation software Method-of-Line (MoL)\cite{LeFloch}. These modes excited in the STO dielectric resonator are experimentally determined with a Vector Network Analyzer (VNA) sensing the modes in transmission $S_{21}$. By matching measured and simulated mode frequencies, the relative permittivity of the crystal has been estimated as $316\pm 0.2$ at room temperature $295\pm1$K.\

Python computer software was used to record the variation of resonance frequency of quasi $TE_{1,1,1}$, quasi $TE_{2,1,1}$ and quasi $TM_{2,1,1}$ modes under continuous cooling from $275~K$ down to liquid helium temperature ($5~K$). This frequency-temperature characteristics from $275~K$ down to $25~K$ is shown in the figure.\ref{STO1}, and also from $75~K$ down to liquid helium temperature ($5~K$) is shown in the separate figure (Fig.\ref{STO2}) for clear visibility of the curves including the trends to be flat at $5~K$. The $TE_{1,1,1}$ mode and $TE_{2,1,1}$ mode resonance frequency $4.555~GHz$ and $6.696~GHz$ decreased down to $456~MHz$ and $678~MHz$ respectively. The $TM_{2,1,1}$ mode resonance frequency $6.539~GHz$ decreased down to about $626~MHz$~$(Fig.\ref{STO1},\ref{STO2})$. Also, there is a little peak in each characteristic curve around $105~K$ $(Fig.\ref{STO1})$, and may imply to the cause of the transition of crystal unit cubic structure to tetragonal structure. The dramatic step-down of dielectric resonance frequency of the observed frequency-temperature characteristic curves indicates an increase of relative permittivity of the crystal up to a figure about $3\times10^4$. The temperature-frequency characteristic curves of different mode families crosses each other due to different polarity of mode, and influence of crystal anisotropy at lower temperatures likely lower than phase transition temperature. Rate of change of the decrease of frequency with respect to temperature becomes more rapid initiating higher curvature around $75~K$ indicates further structural change of crystal unit as an effect of soft mode pre stage. Marked frequency shift $\bigtriangleup\omega'$ and $\bigtriangleup\omega''$ in the figure.\ref{STO1} is likely due to soft-mode. This trends ended to be flat at a temperature lower than $5~K$ due to quantum criticality (Fig.\ref{STO2}) as explained by Rowley et al.\cite{QCriticality}. This frequency-temperature characteristics of $TE_{2,1,1}$ mode individually as a single mode is given in the figure.\ref{STO5} with log scale of temperature.\

\subsubsection{$SrTiO_3$ crystal phase transitions:}

Generally two-level system of Gibbs free energy minima regarding spontaneous polarization $\pm P_s$ is generated in the ferroelectric phase. And $P_s$ switches between these two energy minima as $+P_s$ and $-P_s$ ($Fig.\ref{STO}$) with travelling microwave in the crystal lattice\cite{FE-Swicth,FE-SwitchStraLim}. At the ferroelectric phase at a temperature $T<T_\circ$, this non-zero polarization is usually formed in terms of a softening of an optical mode at the Brillion zone center, which produces the separation of positive and negative charge centers. The imaginary parts of the complex permittivity reflects delayed responses under corresponding external microwave stimuli, and minimum loss of the operating quasi $TE_{m,1,1}$ and quasi $TM_{m,1,1}$ microwave modes are occurred when $P_s = 0$ at the phase transition temperature\cite{LossFE}. Obviously, the FE transition happened at a temperature according to Ginzburg-Landau theory where ${\frac{\partial \phi}{\partial P(T)}}=0$ with respect to parameter ${P(T)}$ of the phenomena\cite{Landau,SDriven}.\

\begin{figure}[t!]
\centering
\includegraphics[ width=0.5\textwidth]{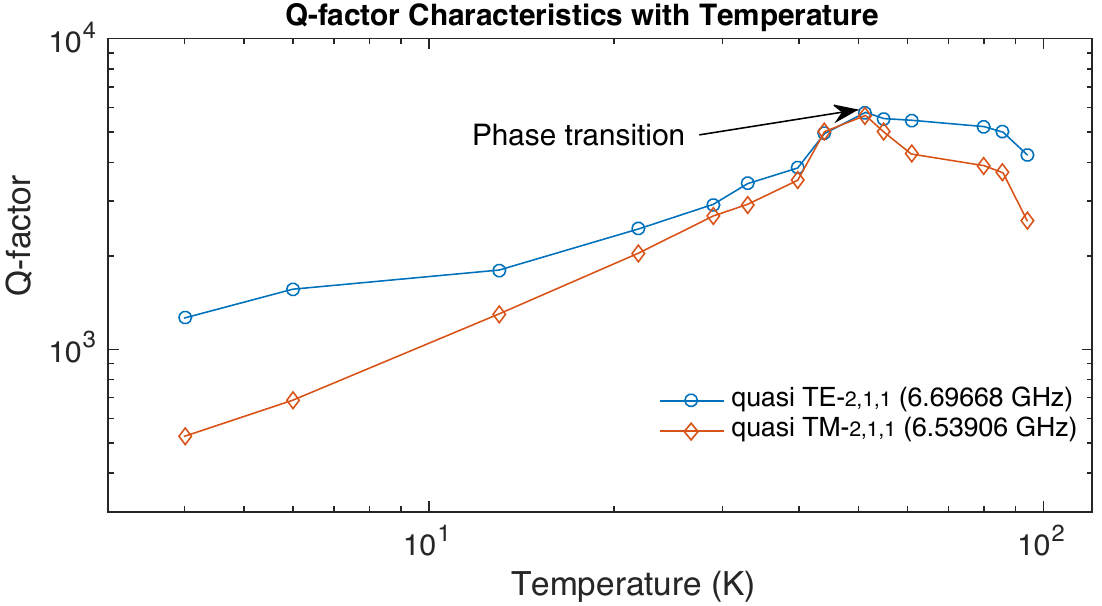}
\caption[Q-factor characteristics with respect to temperature]{\label{STO3} Q-factor characteristics with respect to temperature cooling continuously down to $4~K$. Characteristic curve shows FE phase transition at the point of Q-factor maxima.}
\end{figure}

\begin{figure}[t!]
\centering
\includegraphics[width=0.5\textwidth]{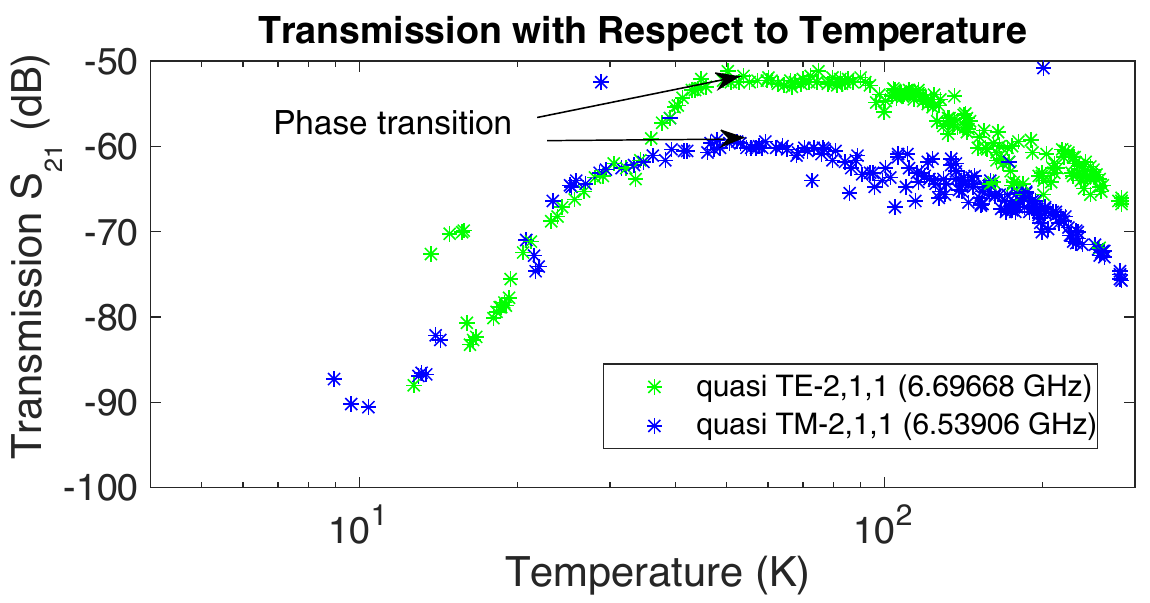}
\caption[Transmission $S_{21}$characteristics with respect to temperature ]{\label{STO4} Transmission $S_{21}$characteristics with respect to temperature cooling continuously down to $4~K$ shows FE phase transition. Modes ar dispersed at quantum criticality state after $4~K$.}
\end{figure}

Rowley et al suggested the FE transition of $SrTiO_3$ as higher than $50~K$ and observed quantum criticality down to $3~K$ temperature\cite{QCriticality}. Our measured FE transition temperature is close to $51~K$ indicating a slight discontinuity in the derivative of $(S_{21})$ near the maxima-point of transmission of the curves as shown in the $Fig.\ref{STO4}$. The flat region around transition temperature  is due to lower rate of change of transmission intensity. Also, the FE phase transition temperature is confirmed showing the near highest $Q$-factor of the $TE_{2,1,1}$, and $TM_{2,1,1}$ modes in the characteristic curves approximately at $51~K$ (Fig.\ref{STO3}). We suggest that the crystal goes under a transition by crystal unit cell structural change from cubic into tetragonal shape at temperatures higher than $51~K$. This transition temperatures may be identified to be about $105~K$ according to the observed noisy tilt in the transmission $(S_{21})$ characteristics with respect to temperature $(Fig.\ref{STO4})$, and similarly a tilt in the curve of frequency-temperature characteristics are observed (Fig.\ref{STO1}). Plausibly, the noisy tilt is the cause of geometric frustration of the crystal unit at the beginning of cubic to tetragonal structural transition, where atoms tend to stick to non-trivial positions on a regular crystal lattice conflicting inter-atomic forces leading to different structures\cite{STO-trivialPos}. Hence, it allows direct imaging of local arrangement in the crystalline lattice with atomic resolution. However, due to the non-trivial influence of thermal diffuse scattering, a detailed examination of the image comparison shows that transmission noise level of $TM_{2,1,1}$ mode is higher or different in temperature scale than $TE_{2,1,1}$ mode ($Fig.\ref{STO4}$). The frequency shifts $\bigtriangleup\omega'$ and $\bigtriangleup\omega''$ in $Fig.\ref{STO1}$ due to soft mode is a measure of the depth of double-well potential, and the strength of spontaneous polarization ($P_s$). The coefficient $\alpha_\circ (T-T_\circ)$ of free energy equation (equation in the caption of Fig.\ref{STO}) is the measure of strength of $P_s$ and the depth of the double-well potential forming a two-level system (Fig.\ref{STO}). The frequency shifts $\bigtriangleup\omega'$ and $\bigtriangleup\omega''$ in $Fig.\ref{STO1}$ is a measure of the square of the  imaginary frequency $\tilde{\omega}_q^2$ (see equation-\eqref{eq:Freeenergy9}) of temperature dependent soft-mode\cite{DisplaciveFE}.\

\begin{figure}[h!]
\centering
\includegraphics[width=0.5\textwidth]{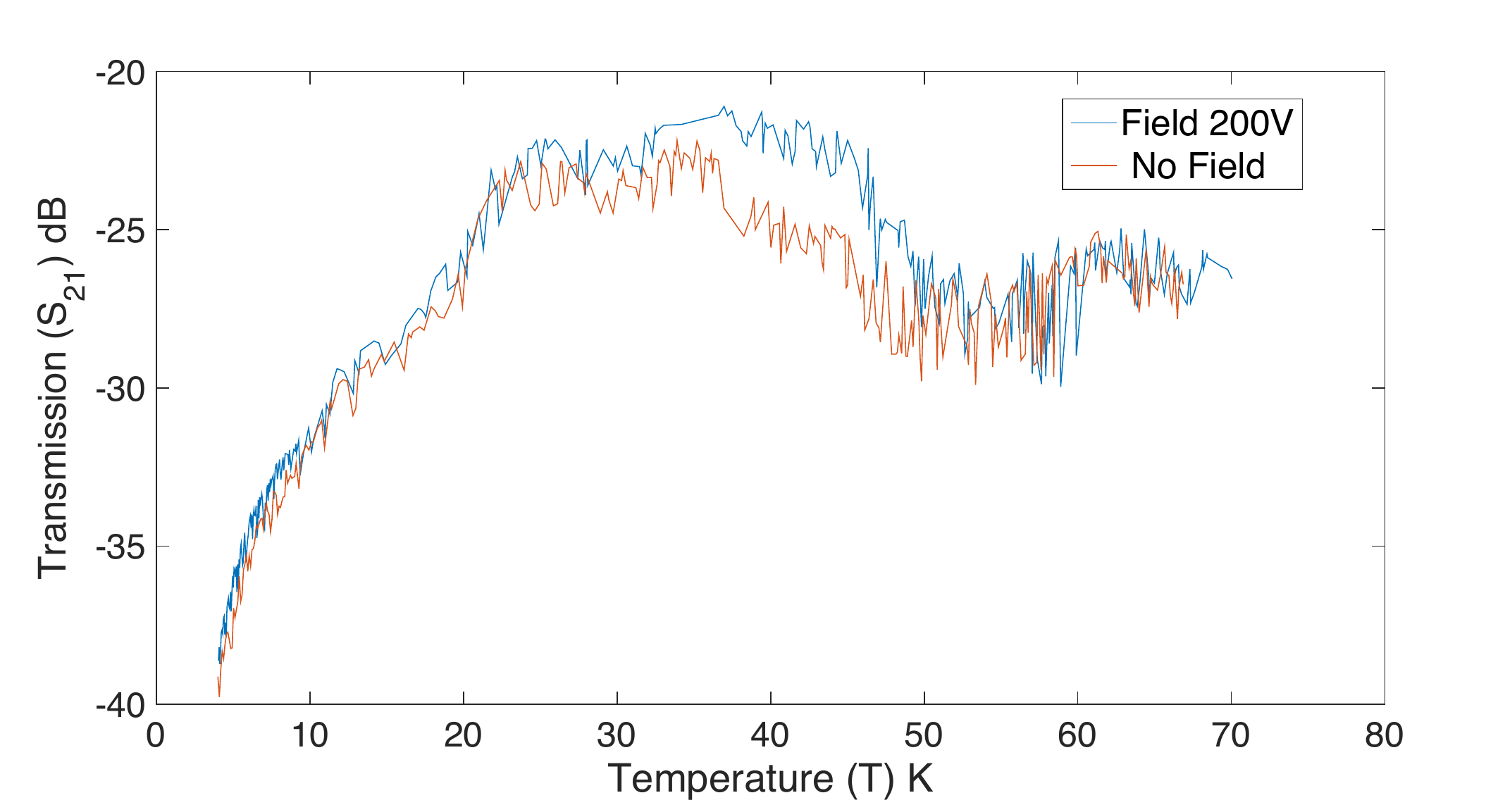}
\caption[Hysteresis characteristics rgarding transmission with respect to temperature]{\label{STO6} $TE_{2,1,1}$ mode transmission $S_{21}$ characteristics with respect to temperature in continuous cooling down to $4~K$ under DC electric field 200 V forms loop (after 51 K) likely FE-hysteresis due to soft mode spontaneous polarization $P_s$.}
\end{figure}

The FE phase transition is significantly indicated showing a loop in the transmission characteristic curve of $TE_{2,1,1}$ mode under external DC electric field 200 V along crystal axis and without this external field in cooling from room temperature down to $4~K$ (Fig.\ref{STO6}). Apparently, the soft-mode produced spontaneous polarization is influenced by the external electric field which is revealed by showing higher transmission intensity of the $TE_{2,1,1}$ mode in the region of soft-mode producing temperature $T\leq 51~K$ of this incipient FE crystal. Such a non-zero spontaneous polarization as a nonlinear polarization forms a $P_s-E_{RF}$ hysteresis loops showing the variation of dielectric polarization with RF-field and the polarization switching behaviour. In principle, formed $P_s-E_{RF}$ loop can  influence $P_s$ resulting to remnant polarization ($P_r$) and $E_{RF}$ to coercive field ($E_c$) in a rate of the cycle of switching with microwave mode. Microwave power dependency was apparently low as predicted for this incipient ferroelectric crystal. The influence of microwave power becomes accountable at lower temperature likely less than $10~K$, where transmission was disappeared after $8~K$ at lower microwave power ($Fig.\ref{STO4}$) and the transmission was observed down to $4~K$ providing comparatively higher microwave power ($Fig.\ref{STO6}$), in the FE phase structural anisotropy state.\

\subsection{Conclusion}
 
 Application of quasi $TE_{m,1,1}$ and quasi $TM_{m,1,1}$ microwave modes provide an opportunity to measure the transition temperature of FE two-level system in a significant process, and an effective tool to diagnose microscopic state of crystal lattice. Basing on statistical mechanics\cite{Kramers}, in an average of thermal fluctuation this state exist as a PE phase with $P_s =0$. This is a strong agreement with FE free energy minimum state requirement. Highly sensitive quasi $TE_{m,1,1}$ mode and quasi $TM_{m,1,1}$ mode probing enables us to monitor such a FE transition like $SrTiO_3$ where mode is not scattered in the depth of FE potential wells.
  
Apparently, Q-factor is showing phase transition temperature lower than the observed soft-mode frequency shift temperature in the frequency-temperature characteristic curves $(Fig.\ref{STO1}$ and $\ref{STO2})$. Mani et al claimed similar fact as dynamical effect in a typical order-disorder mechanism\cite{DynaStatFE}. 

\begin{acknowledgements}
This work was supported by the Australian Research Council Grant No. CE170100009. The authors would like to thanks technical staff Mr. Steve Osborne for support in making copper cavity.
\end{acknowledgements}

\section{References}


\begin{thebibliography}{41}%
\makeatletter
\providecommand \@ifxundefined [1]{%
 \@ifx{#1\undefined}
}%
\providecommand \@ifnum [1]{%
 \ifnum #1\expandafter \@firstoftwo
 \else \expandafter \@secondoftwo
 \fi
}%
\providecommand \@ifx [1]{%
 \ifx #1\expandafter \@firstoftwo
 \else \expandafter \@secondoftwo
 \fi
}%
\providecommand \natexlab [1]{#1}%
\providecommand \enquote  [1]{``#1''}%
\providecommand \bibnamefont  [1]{#1}%
\providecommand \bibfnamefont [1]{#1}%
\providecommand \citenamefont [1]{#1}%
\providecommand \href@noop [0]{\@secondoftwo}%
\providecommand \href [0]{\begingroup \@sanitize@url \@href}%
\providecommand \@href[1]{\@@startlink{#1}\@@href}%
\providecommand \@@href[1]{\endgroup#1\@@endlink}%
\providecommand \@sanitize@url [0]{\catcode `\\12\catcode `\$12\catcode
  `\&12\catcode `\#12\catcode `\^12\catcode `\_12\catcode `\%12\relax}%
\providecommand \@@startlink[1]{}%
\providecommand \@@endlink[0]{}%
\providecommand \url  [0]{\begingroup\@sanitize@url \@url }%
\providecommand \@url [1]{\endgroup\@href {#1}{\urlprefix }}%
\providecommand \urlprefix  [0]{URL }%
\providecommand \Eprint [0]{\href }%
\providecommand \doibase [0]{http://dx.doi.org/}%
\providecommand \selectlanguage [0]{\@gobble}%
\providecommand \bibinfo  [0]{\@secondoftwo}%
\providecommand \bibfield  [0]{\@secondoftwo}%
\providecommand \translation [1]{[#1]}%
\providecommand \BibitemOpen [0]{}%
\providecommand \bibitemStop [0]{}%
\providecommand \bibitemNoStop [0]{.\EOS\space}%
\providecommand \EOS [0]{\spacefactor3000\relax}%
\providecommand \BibitemShut  [1]{\csname bibitem#1\endcsname}%
\let\auto@bib@innerbib\@empty
\bibitem [{\citenamefont {BussmannHolder}(2012)}]{Polariz}%
  \BibitemOpen
  \bibfield  {author} {\bibinfo {author} {\bibfnamefont {A.}~\bibnamefont
  {BussmannHolder}},\ }\href@noop {} {\bibfield  {journal} {\bibinfo  {journal}
  {J. Physics Condensed Matter}\ }\textbf {\bibinfo {volume} {24}},\ \bibinfo
  {pages} {273202} (\bibinfo {year} {2012})}\BibitemShut {NoStop}%
\bibitem [{\citenamefont {Dove}(1997)}]{DisplaciveFE}%
  \BibitemOpen
  \bibfield  {author} {\bibinfo {author} {\bibfnamefont {M.~T.}\ \bibnamefont
  {Dove}},\ }\href@noop {} {\bibfield  {journal} {\bibinfo  {journal} {American
  Minaraligist}\ }\textbf {\bibinfo {volume} {82}},\ \bibinfo {pages} {213}
  (\bibinfo {year} {1997})}\BibitemShut {NoStop}%
\bibitem [{\citenamefont {Gomilsek}(2012)}]{WG}%
  \BibitemOpen
  \bibfield  {author} {\bibinfo {author} {\bibfnamefont {M.}~\bibnamefont
  {Gomilsek}},\ }\href@noop {} {\bibfield  {journal} {\bibinfo  {journal}
  {Seminer: University of Ljubljana, Slovenia.}\ } (\bibinfo {year}
  {2012})}\BibitemShut {NoStop}%
\bibitem [{\citenamefont {Le-Floch}(2009)}]{LeFloch}%
  \BibitemOpen
  \bibfield  {author} {\bibinfo {author} {\bibfnamefont {J.~M.}\ \bibnamefont
  {Le-Floch}},\ }\href@noop {} {\emph {\bibinfo {title} {Thesis: Modelling of
  New high-Q Dielectric Sensors for Metrology Applications}}}\ (\bibinfo
  {publisher} {School of Physics, University of Western Australia, Crawley, WA
  6009},\ \bibinfo {year} {2009})\BibitemShut {NoStop}%
\bibitem [{\citenamefont {Krupka}\ \emph {et~al.}(1999)\citenamefont {Krupka},
  \citenamefont {Derzakowiski}, \citenamefont {Abramowicz}, \citenamefont
  {Tobar},\ and\ \citenamefont {Geyer}}]{Krupka}%
  \BibitemOpen
  \bibfield  {author} {\bibinfo {author} {\bibfnamefont {J.}~\bibnamefont
  {Krupka}}, \bibinfo {author} {\bibfnamefont {K.}~\bibnamefont
  {Derzakowiski}}, \bibinfo {author} {\bibfnamefont {A.}~\bibnamefont
  {Abramowicz}}, \bibinfo {author} {\bibfnamefont {M.~E.}\ \bibnamefont
  {Tobar}}, \ and\ \bibinfo {author} {\bibfnamefont {R.~G.}\ \bibnamefont
  {Geyer}},\ }\href@noop {} {\bibfield  {journal} {\bibinfo  {journal} {Meas.
  Sci. Technol.}\ }\textbf {\bibinfo {volume} {10}},\ \bibinfo {pages} {387}
  (\bibinfo {year} {1999})}\BibitemShut {NoStop}%
\bibitem [{\citenamefont {Krupka}\ and\ \citenamefont
  {Mazierska}(2000)}]{Krupka1}%
  \BibitemOpen
  \bibfield  {author} {\bibinfo {author} {\bibfnamefont {J.}~\bibnamefont
  {Krupka}}\ and\ \bibinfo {author} {\bibfnamefont {J.}~\bibnamefont
  {Mazierska}},\ }\href@noop {} {\bibfield  {journal} {\bibinfo  {journal}
  {IEEE Transactions On Microwave Theory And Techniques}\ }\textbf {\bibinfo
  {volume} {Special Issue}},\ \bibinfo {pages} {1} (\bibinfo {year}
  {2000})}\BibitemShut {NoStop}%
\bibitem [{\citenamefont {Konopka}\ and\ \citenamefont
  {Wolff}(1992)}]{Konopka}%
  \BibitemOpen
  \bibfield  {author} {\bibinfo {author} {\bibfnamefont {J.}~\bibnamefont
  {Konopka}}\ and\ \bibinfo {author} {\bibfnamefont {I.}~\bibnamefont
  {Wolff}},\ }\href@noop {} {\bibfield  {journal} {\bibinfo  {journal} {IEEE
  Transactions On Microwave Theory And Techniques}\ }\textbf {\bibinfo {volume}
  {40}},\ \bibinfo {pages} {1027} (\bibinfo {year} {1992})}\BibitemShut
  {NoStop}%
\bibitem [{\citenamefont {Pajaczkowaska}\ and\ \citenamefont
  {Gloubokov}(1998)}]{Anna}%
  \BibitemOpen
  \bibfield  {author} {\bibinfo {author} {\bibfnamefont {A.}~\bibnamefont
  {Pajaczkowaska}}\ and\ \bibinfo {author} {\bibfnamefont {A.}~\bibnamefont
  {Gloubokov}},\ }\href@noop {} {\bibfield  {journal} {\bibinfo  {journal}
  {Prog. Crystal Growth and Charact.}\ }\textbf {\bibinfo {volume} {36}},\
  \bibinfo {pages} {123} (\bibinfo {year} {1998})}\BibitemShut {NoStop}%
\bibitem [{\citenamefont {Mitchell}, \citenamefont {Chakhmouradian},\ and\
  \citenamefont {Woodward}(2000)}]{STO-Structure}%
  \BibitemOpen
  \bibfield  {author} {\bibinfo {author} {\bibfnamefont {R.~H.}\ \bibnamefont
  {Mitchell}}, \bibinfo {author} {\bibfnamefont {A.~R.}\ \bibnamefont
  {Chakhmouradian}}, \ and\ \bibinfo {author} {\bibfnamefont {P.~M.}\
  \bibnamefont {Woodward}},\ }\href@noop {} {\bibfield  {journal} {\bibinfo
  {journal} {Phys. Chem. Miner. (Germany)}\ }\textbf {\bibinfo {volume} {27}},\
  \bibinfo {pages} {583} (\bibinfo {year} {2000})}\BibitemShut {NoStop}%
\bibitem [{\citenamefont {Brik}\ and\ \citenamefont
  {Avram}(2009)}]{STO-CrystField1}%
  \BibitemOpen
  \bibfield  {author} {\bibinfo {author} {\bibfnamefont {M.~G.}\ \bibnamefont
  {Brik}}\ and\ \bibinfo {author} {\bibfnamefont {N.~M.}\ \bibnamefont
  {Avram}},\ }\href {\doibase 10.1088/0953-8984/21/15/155502} {\bibfield
  {journal} {\bibinfo  {journal} {J. Phys.: Condens. Matter}\ }\textbf
  {\bibinfo {volume} {21}},\ \bibinfo {pages} {155502(9pp.)} (\bibinfo {year}
  {2009})}\BibitemShut {NoStop}%
\bibitem [{\citenamefont {Cohen}(1992)}]{OriginFE}%
  \BibitemOpen
  \bibfield  {author} {\bibinfo {author} {\bibfnamefont {R.~E.}\ \bibnamefont
  {Cohen}},\ }\href {\doibase 10.1038/358136a0} {\bibfield  {journal} {\bibinfo
   {journal} {NATURE}\ }\textbf {\bibinfo {volume} {358}},\ \bibinfo {pages}
  {136} (\bibinfo {year} {1992})}\BibitemShut {NoStop}%
\bibitem [{\citenamefont {Hohenberg}\ and\ \citenamefont
  {Krekhov}(2015)}]{Landau}%
  \BibitemOpen
  \bibfield  {author} {\bibinfo {author} {\bibfnamefont {P.~C.}\ \bibnamefont
  {Hohenberg}}\ and\ \bibinfo {author} {\bibfnamefont {A.~P.}\ \bibnamefont
  {Krekhov}},\ }\href@noop {} {\bibfield  {journal} {\bibinfo  {journal}
  {Physics Reports}\ }\textbf {\bibinfo {volume} {572}},\ \bibinfo {pages} {1}
  (\bibinfo {year} {2015})}\BibitemShut {NoStop}%
\bibitem [{\citenamefont {Dove}(1993)}]{DisplaciveFE_Book}%
  \BibitemOpen
  \bibfield  {author} {\bibinfo {author} {\bibfnamefont {M.~T.}\ \bibnamefont
  {Dove}},\ }\href@noop {} {\emph {\bibinfo {title} {Introduction to lattice
  dynamics}}}\ (\bibinfo  {publisher} {Cambridge University Press, Cambridge},\
  \bibinfo {year} {1993})\ p.~\bibinfo {pages} {58}\BibitemShut {NoStop}%
\bibitem [{\citenamefont {Weng}\ \emph {et~al.}(2014)\citenamefont {Weng},
  \citenamefont {Xu}, \citenamefont {Said}, \citenamefont {Leu}, \citenamefont
  {Ding}, \citenamefont {Hong}, \citenamefont {Fang}, \citenamefont {Chou},
  \citenamefont {Bosak}, \citenamefont {Abbamonte}, \citenamefont {Cooper},
  \citenamefont {Fradkin}, \citenamefont {Chang},\ and\ \citenamefont
  {Chiang}}]{PressAntiFerro}%
  \BibitemOpen
  \bibfield  {author} {\bibinfo {author} {\bibfnamefont {S.-C.}\ \bibnamefont
  {Weng}}, \bibinfo {author} {\bibfnamefont {R.}~\bibnamefont {Xu}}, \bibinfo
  {author} {\bibfnamefont {A.~H.}\ \bibnamefont {Said}}, \bibinfo {author}
  {\bibfnamefont {B.~M.}\ \bibnamefont {Leu}}, \bibinfo {author} {\bibfnamefont
  {Y.}~\bibnamefont {Ding}}, \bibinfo {author} {\bibfnamefont {H.}~\bibnamefont
  {Hong}}, \bibinfo {author} {\bibfnamefont {X.}~\bibnamefont {Fang}}, \bibinfo
  {author} {\bibfnamefont {M.~Y.}\ \bibnamefont {Chou}}, \bibinfo {author}
  {\bibfnamefont {A.}~\bibnamefont {Bosak}}, \bibinfo {author} {\bibfnamefont
  {P.}~\bibnamefont {Abbamonte}}, \bibinfo {author} {\bibfnamefont {S.~L.}\
  \bibnamefont {Cooper}}, \bibinfo {author} {\bibfnamefont {E.}~\bibnamefont
  {Fradkin}}, \bibinfo {author} {\bibfnamefont {S.-L.}\ \bibnamefont {Chang}},
  \ and\ \bibinfo {author} {\bibfnamefont {T.-C.}\ \bibnamefont {Chiang}},\
  }\href {\doibase 10.1209/0295-5075/107/36006} {\bibfield  {journal} {\bibinfo
   {journal} {EPL}\ }\textbf {\bibinfo {volume} {107}},\ \bibinfo {pages}
  {36006 pp.5} (\bibinfo {year} {2014})}\BibitemShut {NoStop}%
\bibitem [{\citenamefont {Salje}\ \emph {et~al.}(2014)\citenamefont {Salje},
  \citenamefont {Guennou}, \citenamefont {Bouvier}, \citenamefont {Carpenter},\
  and\ \citenamefont {Kreisel}}]{PressAntiFerro1}%
  \BibitemOpen
  \bibfield  {author} {\bibinfo {author} {\bibfnamefont {E.~K.~H.}\
  \bibnamefont {Salje}}, \bibinfo {author} {\bibfnamefont {M.}~\bibnamefont
  {Guennou}}, \bibinfo {author} {\bibfnamefont {P.}~\bibnamefont {Bouvier}},
  \bibinfo {author} {\bibfnamefont {M.~A.}\ \bibnamefont {Carpenter}}, \ and\
  \bibinfo {author} {\bibfnamefont {J.}~\bibnamefont {Kreisel}},\ }\href
  {\doibase 10.1088/0953-8984/23/27/275901} {\bibfield  {journal} {\bibinfo
  {journal} {J. Phys.: Condens. Matter}\ }\textbf {\bibinfo {volume} {23}},\
  \bibinfo {pages} {275901 pp.5} (\bibinfo {year} {2014})}\BibitemShut
  {NoStop}%
\bibitem [{\citenamefont {M$\ddot{u}$ller}\ and\ \citenamefont
  {Burkard}(1979)}]{MullerQuanPE}%
  \BibitemOpen
  \bibfield  {author} {\bibinfo {author} {\bibfnamefont {K.~A.}\ \bibnamefont
  {M$\ddot{u}$ller}}\ and\ \bibinfo {author} {\bibfnamefont {H.}~\bibnamefont
  {Burkard}},\ }\href@noop {} {\bibfield  {journal} {\bibinfo  {journal}
  {Physical Rview B}\ }\textbf {\bibinfo {volume} {19}},\ \bibinfo {pages}
  {3595} (\bibinfo {year} {1979})}\BibitemShut {NoStop}%
\bibitem [{\citenamefont {Haeni}\ \emph {et~al.}(2004)\citenamefont {Haeni},
  \citenamefont {Irvin}, \citenamefont {Chang}, \citenamefont {Uecker},
  \citenamefont {Reiche}, \citenamefont {Li}, \citenamefont {Choudhury},
  \citenamefont {Tian}, \citenamefont {Hawley}, \citenamefont {Craigo},
  \citenamefont {Tagantsev}, \citenamefont {Pan}, \citenamefont {Streiffer},
  \citenamefont {Chen}, \citenamefont {Kirchoefer}, \citenamefont {Levy},\ and\
  \citenamefont {Schlom}}]{RoomTemFE}%
  \BibitemOpen
  \bibfield  {author} {\bibinfo {author} {\bibfnamefont {J.~H.}\ \bibnamefont
  {Haeni}}, \bibinfo {author} {\bibfnamefont {P.}~\bibnamefont {Irvin}},
  \bibinfo {author} {\bibfnamefont {W.}~\bibnamefont {Chang}}, \bibinfo
  {author} {\bibfnamefont {R.}~\bibnamefont {Uecker}}, \bibinfo {author}
  {\bibfnamefont {P.}~\bibnamefont {Reiche}}, \bibinfo {author} {\bibfnamefont
  {Y.~L.}\ \bibnamefont {Li}}, \bibinfo {author} {\bibfnamefont
  {S.}~\bibnamefont {Choudhury}}, \bibinfo {author} {\bibfnamefont
  {W.}~\bibnamefont {Tian}}, \bibinfo {author} {\bibfnamefont {M.~E.}\
  \bibnamefont {Hawley}}, \bibinfo {author} {\bibfnamefont {B.}~\bibnamefont
  {Craigo}}, \bibinfo {author} {\bibfnamefont {A.~K.}\ \bibnamefont
  {Tagantsev}}, \bibinfo {author} {\bibfnamefont {X.~Q.}\ \bibnamefont {Pan}},
  \bibinfo {author} {\bibfnamefont {S.~K.}\ \bibnamefont {Streiffer}}, \bibinfo
  {author} {\bibfnamefont {L.~Q.}\ \bibnamefont {Chen}}, \bibinfo {author}
  {\bibfnamefont {S.~W.}\ \bibnamefont {Kirchoefer}}, \bibinfo {author}
  {\bibfnamefont {J.}~\bibnamefont {Levy}}, \ and\ \bibinfo {author}
  {\bibfnamefont {D.~G.}\ \bibnamefont {Schlom}},\ }\href@noop {} {\bibfield
  {journal} {\bibinfo  {journal} {NATURE}\ }\textbf {\bibinfo {volume} {430}},\
  \bibinfo {pages} {758} (\bibinfo {year} {2004})}\BibitemShut {NoStop}%
\bibitem [{\citenamefont {Jang}\ \emph {et~al.}(2010)\citenamefont {Jang},
  \citenamefont {Kumar}, \citenamefont {Denev}, \citenamefont {Biegalski},
  \citenamefont {Maksymovych}, \citenamefont {Bark}, \citenamefont {Nelson},
  \citenamefont {Folkman}, \citenamefont {Baek}, \citenamefont {Balke},
  \citenamefont {Brooks}, \citenamefont {Tenne}, \citenamefont {Schlom},
  \citenamefont {Chen}, \citenamefont {Pan}, \citenamefont {Kalinin},
  \citenamefont {Gopalan},\ and\ \citenamefont {Eom}}]{ComparisonFE}%
  \BibitemOpen
  \bibfield  {author} {\bibinfo {author} {\bibfnamefont {ï.~W.}\ \bibnamefont
  {Jang}}, \bibinfo {author} {\bibfnamefont {A.}~\bibnamefont {Kumar}},
  \bibinfo {author} {\bibfnamefont {S.}~\bibnamefont {Denev}}, \bibinfo
  {author} {\bibfnamefont {M.~D.}\ \bibnamefont {Biegalski}}, \bibinfo {author}
  {\bibfnamefont {P.}~\bibnamefont {Maksymovych}}, \bibinfo {author}
  {\bibfnamefont {C.~W.}\ \bibnamefont {Bark}}, \bibinfo {author}
  {\bibfnamefont {C.~T.}\ \bibnamefont {Nelson}}, \bibinfo {author}
  {\bibfnamefont {C.~M.}\ \bibnamefont {Folkman}}, \bibinfo {author}
  {\bibfnamefont {S.~H.}\ \bibnamefont {Baek}}, \bibinfo {author}
  {\bibfnamefont {N.}~\bibnamefont {Balke}}, \bibinfo {author} {\bibfnamefont
  {C.~M.}\ \bibnamefont {Brooks}}, \bibinfo {author} {\bibfnamefont {D.~A.}\
  \bibnamefont {Tenne}}, \bibinfo {author} {\bibfnamefont {D.~G.}\ \bibnamefont
  {Schlom}}, \bibinfo {author} {\bibfnamefont {L.~Q.}\ \bibnamefont {Chen}},
  \bibinfo {author} {\bibfnamefont {X.~Q.}\ \bibnamefont {Pan}}, \bibinfo
  {author} {\bibfnamefont {S.~V.}\ \bibnamefont {Kalinin}}, \bibinfo {author}
  {\bibfnamefont {V.}~\bibnamefont {Gopalan}}, \ and\ \bibinfo {author}
  {\bibfnamefont {C.~B.}\ \bibnamefont {Eom}},\ }\href {\doibase
  10.1103/PhysRevLett.104.197601} {\bibfield  {journal} {\bibinfo  {journal}
  {Physical Reviw Letters}\ }\textbf {\bibinfo {volume} {104}},\ \bibinfo
  {pages} {197601 pp.4} (\bibinfo {year} {2010})}\BibitemShut {NoStop}%
\bibitem [{\citenamefont {Cohen}\ and\ \citenamefont
  {Krakauer}(1990)}]{STO-BTO-lattice}%
  \BibitemOpen
  \bibfield  {author} {\bibinfo {author} {\bibfnamefont {R.~E.}\ \bibnamefont
  {Cohen}}\ and\ \bibinfo {author} {\bibfnamefont {H.}~\bibnamefont
  {Krakauer}},\ }\href {\doibase https://doi.org/10.1103/PhysRevB.42.6416}
  {\bibfield  {journal} {\bibinfo  {journal} {Phys. Rev. B}\ }\textbf {\bibinfo
  {volume} {42}},\ \bibinfo {pages} {6416} (\bibinfo {year}
  {1990})}\BibitemShut {NoStop}%
\bibitem [{\citenamefont {Benedek}\ and\ \citenamefont
  {Fennie}(2013)}]{STO-FewFE}%
  \BibitemOpen
  \bibfield  {author} {\bibinfo {author} {\bibfnamefont {N.~A.}\ \bibnamefont
  {Benedek}}\ and\ \bibinfo {author} {\bibfnamefont {C.~J.}\ \bibnamefont
  {Fennie}},\ }\href {\doibase 10.1021/jp402046t} {\bibfield  {journal}
  {\bibinfo  {journal} {J. Phys. Chem. C}\ }\textbf {\bibinfo {volume} {117}},\
  \bibinfo {pages} {13339} (\bibinfo {year} {2013})}\BibitemShut {NoStop}%
\bibitem [{\citenamefont {Pierangeli}\ \emph {et~al.}(2016)\citenamefont
  {Pierangeli}, \citenamefont {Ferraro}, \citenamefont {Mei}, \citenamefont
  {Domenico}, \citenamefont {de~Oliveira}, \citenamefont {Agranat},\ and\
  \citenamefont {DelRe}}]{SpontPola}%
  \BibitemOpen
  \bibfield  {author} {\bibinfo {author} {\bibfnamefont {D.}~\bibnamefont
  {Pierangeli}}, \bibinfo {author} {\bibfnamefont {M.}~\bibnamefont {Ferraro}},
  \bibinfo {author} {\bibfnamefont {F.~D.}\ \bibnamefont {Mei}}, \bibinfo
  {author} {\bibfnamefont {G.~D.}\ \bibnamefont {Domenico}}, \bibinfo {author}
  {\bibfnamefont {C.~E.~M.}\ \bibnamefont {de~Oliveira}}, \bibinfo {author}
  {\bibfnamefont {A.~J.}\ \bibnamefont {Agranat}}, \ and\ \bibinfo {author}
  {\bibfnamefont {E.}~\bibnamefont {DelRe}},\ }\href {\doibase
  10.1038/ncomms10674} {\bibfield  {journal} {\bibinfo  {journal} {Nature
  Communications}\ }\textbf {\bibinfo {volume} {7}},\ \bibinfo {pages} {1}
  (\bibinfo {year} {2016})}\BibitemShut {NoStop}%
\bibitem [{\citenamefont {Piskunov}\ \emph {et~al.}(2004)\citenamefont
  {Piskunov}, \citenamefont {Heifets}, \citenamefont {Eglitis},\ and\
  \citenamefont {Borstel}}]{STO-Struc}%
  \BibitemOpen
  \bibfield  {author} {\bibinfo {author} {\bibfnamefont {S.}~\bibnamefont
  {Piskunov}}, \bibinfo {author} {\bibfnamefont {E.}~\bibnamefont {Heifets}},
  \bibinfo {author} {\bibfnamefont {R.}~\bibnamefont {Eglitis}}, \ and\
  \bibinfo {author} {\bibfnamefont {G.}~\bibnamefont {Borstel}},\ }\href@noop
  {} {\bibfield  {journal} {\bibinfo  {journal} {Computational Materials
  Science}\ }\textbf {\bibinfo {volume} {29}},\ \bibinfo {pages} {165}
  (\bibinfo {year} {2004})}\BibitemShut {NoStop}%
\bibitem [{\citenamefont {Liu}\ \emph {et~al.}(2015)\citenamefont {Liu},
  \citenamefont {Zhang}, \citenamefont {Jiang},\ and\ \citenamefont
  {Cao}}]{LossFE}%
  \BibitemOpen
  \bibfield  {author} {\bibinfo {author} {\bibfnamefont {G.}~\bibnamefont
  {Liu}}, \bibinfo {author} {\bibfnamefont {S.}~\bibnamefont {Zhang}}, \bibinfo
  {author} {\bibfnamefont {W.}~\bibnamefont {Jiang}}, \ and\ \bibinfo {author}
  {\bibfnamefont {W.}~\bibnamefont {Cao}},\ }\href {\doibase
  10.1016/j.mser.2015.01.002} {\bibfield  {journal} {\bibinfo  {journal} {Mater
  Sci Eng R Rep.}\ }\textbf {\bibinfo {volume} {89}},\ \bibinfo {pages} {1}
  (\bibinfo {year} {2015})}\BibitemShut {NoStop}%
\bibitem [{\citenamefont {Thome}, \citenamefont {Silva},\ and\ \citenamefont
  {Alvim}(1997)}]{PseudoFE}%
  \BibitemOpen
  \bibfield  {author} {\bibinfo {author} {\bibfnamefont {Z.}~\bibnamefont
  {Thome}}, \bibinfo {author} {\bibfnamefont {F.~D.}\ \bibnamefont {Silva}}, \
  and\ \bibinfo {author} {\bibfnamefont {A.}~\bibnamefont {Alvim}},\
  }\href@noop {} {\bibfield  {journal} {\bibinfo  {journal} {Ann. Nucl.
  Energy}\ }\textbf {\bibinfo {volume} {24}},\ \bibinfo {pages} {955} (\bibinfo
  {year} {1997})}\BibitemShut {NoStop}%
\bibitem [{\citenamefont {Murzin}, \citenamefont {Pasynkov},\ and\
  \citenamefont {Solov'ev}(1968)}]{SoftMode}%
  \BibitemOpen
  \bibfield  {author} {\bibinfo {author} {\bibfnamefont {V.~N.}\ \bibnamefont
  {Murzin}}, \bibinfo {author} {\bibfnamefont {R.}~\bibnamefont {Pasynkov}}, \
  and\ \bibinfo {author} {\bibfnamefont {S.}~\bibnamefont {Solov'ev}},\
  }\href@noop {} {\bibfield  {journal} {\bibinfo  {journal} {Soviet Physics
  Uspekhi}\ }\textbf {\bibinfo {volume} {10}},\ \bibinfo {pages} {4} (\bibinfo
  {year} {1968})}\BibitemShut {NoStop}%
\bibitem [{\citenamefont {AHN}\ and\ \citenamefont {KIM}(2005)}]{Hardmode}%
  \BibitemOpen
  \bibfield  {author} {\bibinfo {author} {\bibfnamefont {S.-J.}\ \bibnamefont
  {AHN}}\ and\ \bibinfo {author} {\bibfnamefont {J.-J.}\ \bibnamefont {KIM}},\
  }\href@noop {} {\bibfield  {journal} {\bibinfo  {journal} {Japanese Journal
  of Applied Physics}\ }\textbf {\bibinfo {volume} {44}},\ \bibinfo {pages}
  {5122} (\bibinfo {year} {2005})}\BibitemShut {NoStop}%
\bibitem [{\citenamefont {Bokov}\ and\ \citenamefont {Ye}(2006)}]{Relaxor}%
  \BibitemOpen
  \bibfield  {author} {\bibinfo {author} {\bibfnamefont {A.~A.}\ \bibnamefont
  {Bokov}}\ and\ \bibinfo {author} {\bibfnamefont {Z.~G.}\ \bibnamefont {Ye}},\
  }\href@noop {} {\bibfield  {journal} {\bibinfo  {journal} {Journal of
  Materials Science}\ }\textbf {\bibinfo {volume} {41}},\ \bibinfo {pages} {31}
  (\bibinfo {year} {2006})}\BibitemShut {NoStop}%
\bibitem [{\citenamefont {Trainer}(2000)}]{Curie}%
  \BibitemOpen
  \bibfield  {author} {\bibinfo {author} {\bibfnamefont {M.}~\bibnamefont
  {Trainer}},\ }\href@noop {} {\bibfield  {journal} {\bibinfo  {journal}
  {Eur.J.Phys.}\ }\textbf {\bibinfo {volume} {21}},\ \bibinfo {pages} {459}
  (\bibinfo {year} {2000})}\BibitemShut {NoStop}%
\bibitem [{\citenamefont {Ferrarelli}\ \emph {et~al.}(2010)\citenamefont
  {Ferrarelli}, \citenamefont {Nuzhnyy}, \citenamefont {Scinclair},\ and\
  \citenamefont {Kamba}}]{Soft}%
  \BibitemOpen
  \bibfield  {author} {\bibinfo {author} {\bibfnamefont {M.~C.}\ \bibnamefont
  {Ferrarelli}}, \bibinfo {author} {\bibfnamefont {D.}~\bibnamefont {Nuzhnyy}},
  \bibinfo {author} {\bibfnamefont {D.~C.}\ \bibnamefont {Scinclair}}, \ and\
  \bibinfo {author} {\bibfnamefont {S.}~\bibnamefont {Kamba}},\ }\href@noop {}
  {\bibfield  {journal} {\bibinfo  {journal} {Physics Review}\ }\textbf
  {\bibinfo {volume} {B81}},\ \bibinfo {pages} {224112} (\bibinfo {year}
  {2010})}\BibitemShut {NoStop}%
\bibitem [{\citenamefont {Mccash}(2014)}]{SDriven}%
  \BibitemOpen
  \bibfield  {author} {\bibinfo {author} {\bibfnamefont {K.}~\bibnamefont
  {Mccash}},\ }\href@noop {} {\bibfield  {journal} {\bibinfo  {journal}
  {Thesis: Department of Physics, University of South Florida}\ } (\bibinfo
  {year} {2014})}\BibitemShut {NoStop}%
\bibitem [{\citenamefont {Perry}\ \emph {et~al.}(1989)\citenamefont {Perry},
  \citenamefont {Currat}, \citenamefont {Buhay}, \citenamefont {Migoni},
  \citenamefont {Stirling},\ and\ \citenamefont {Axe}}]{STO-CoreShell}%
  \BibitemOpen
  \bibfield  {author} {\bibinfo {author} {\bibfnamefont {C.~H.}\ \bibnamefont
  {Perry}}, \bibinfo {author} {\bibfnamefont {R.}~\bibnamefont {Currat}},
  \bibinfo {author} {\bibfnamefont {H.}~\bibnamefont {Buhay}}, \bibinfo
  {author} {\bibfnamefont {R.~M.}\ \bibnamefont {Migoni}}, \bibinfo {author}
  {\bibfnamefont {W.~G.}\ \bibnamefont {Stirling}}, \ and\ \bibinfo {author}
  {\bibfnamefont {J.~D.}\ \bibnamefont {Axe}},\ }\href {\doibase
  doi.org/10.1103/PhysRevB.39.8666} {\bibfield  {journal} {\bibinfo  {journal}
  {Phys. Rev. B}\ }\textbf {\bibinfo {volume} {39}},\ \bibinfo {pages} {8666}
  (\bibinfo {year} {1989})}\BibitemShut {NoStop}%
\bibitem [{\citenamefont {Kubo}(2015)}]{Fluctu}%
  \BibitemOpen
  \bibfield  {author} {\bibinfo {author} {\bibfnamefont {R.}~\bibnamefont
  {Kubo}},\ }\href@noop {} {\bibfield  {journal} {\bibinfo  {journal}
  {Department of Physics, University of Tokyo, Japan}\ } (\bibinfo {year}
  {2015})}\BibitemShut {NoStop}%
\bibitem [{\citenamefont {Bussmann}\ \emph {et~al.}(1980)\citenamefont
  {Bussmann}, \citenamefont {Bilz}, \citenamefont {Roenspiess},\ and\
  \citenamefont {Schwarz}}]{STO-QuOxyPola}%
  \BibitemOpen
  \bibfield  {author} {\bibinfo {author} {\bibfnamefont {A.}~\bibnamefont
  {Bussmann}}, \bibinfo {author} {\bibfnamefont {H.}~\bibnamefont {Bilz}},
  \bibinfo {author} {\bibfnamefont {R.}~\bibnamefont {Roenspiess}}, \ and\
  \bibinfo {author} {\bibfnamefont {K.}~\bibnamefont {Schwarz}},\ }\href
  {\doibase https://doi.org/10.1080/00150198008207016} {\bibfield  {journal}
  {\bibinfo  {journal} {Ferroelectrics}\ }\textbf {\bibinfo {volume} {25}},\
  \bibinfo {pages} {343} (\bibinfo {year} {1980})}\BibitemShut {NoStop}%
\bibitem [{\citenamefont {Watson}(1958)}]{STO-Watson}%
  \BibitemOpen
  \bibfield  {author} {\bibinfo {author} {\bibfnamefont {R.~E.}\ \bibnamefont
  {Watson}},\ }\href {\doibase https://doi.org/10.1103/PhysRev.111.1108}
  {\bibfield  {journal} {\bibinfo  {journal} {Phys. Rev.}\ }\textbf {\bibinfo
  {volume} {111}},\ \bibinfo {pages} {1108} (\bibinfo {year}
  {1958})}\BibitemShut {NoStop}%
\bibitem [{\citenamefont {Blink}\ and\ \citenamefont
  {Zeks}(1974)}]{DisplaciveFE_Book1}%
  \BibitemOpen
  \bibfield  {author} {\bibinfo {author} {\bibfnamefont {R.}~\bibnamefont
  {Blink}}\ and\ \bibinfo {author} {\bibfnamefont {B.}~\bibnamefont {Zeks}},\
  }\href@noop {} {\emph {\bibinfo {title} {Soft modes in ferroelectrics and
  antiferroelectrics}}}\ (\bibinfo  {publisher} {American Elsevier Pub. Co.,
  New York},\ \bibinfo {year} {1974})\BibitemShut {NoStop}%
\bibitem [{\citenamefont {Rowley}\ \emph {et~al.}(2014)\citenamefont {Rowley},
  \citenamefont {Spalek}, \citenamefont {Smith}, \citenamefont {Dean},
  \citenamefont {Itoh}, \citenamefont {Scott}, \citenamefont {Lonzarich},\ and\
  \citenamefont {Saxena}}]{QCriticality}%
  \BibitemOpen
  \bibfield  {author} {\bibinfo {author} {\bibfnamefont {S.~E.}\ \bibnamefont
  {Rowley}}, \bibinfo {author} {\bibfnamefont {L.~J.}\ \bibnamefont {Spalek}},
  \bibinfo {author} {\bibfnamefont {R.~P.}\ \bibnamefont {Smith}}, \bibinfo
  {author} {\bibfnamefont {M.~P.~M.}\ \bibnamefont {Dean}}, \bibinfo {author}
  {\bibfnamefont {M.}~\bibnamefont {Itoh}}, \bibinfo {author} {\bibfnamefont
  {J.~F.}\ \bibnamefont {Scott}}, \bibinfo {author} {\bibfnamefont {G.~G.}\
  \bibnamefont {Lonzarich}}, \ and\ \bibinfo {author} {\bibfnamefont {S.~S.}\
  \bibnamefont {Saxena}},\ }\href@noop {} {\bibfield  {journal} {\bibinfo
  {journal} {Nature Physics}\ }\textbf {\bibinfo {volume} {10}},\ \bibinfo
  {pages} {367} (\bibinfo {year} {2014})}\BibitemShut {NoStop}%
\bibitem [{\citenamefont {Nelson}\ \emph {et~al.}(2011)\citenamefont {Nelson},
  \citenamefont {Gao}, \citenamefont {Jokisaari}, \citenamefont {Heikes},
  \citenamefont {Adamo}, \citenamefont {Melville}, \citenamefont {Baek},
  \citenamefont {Folkman}, \citenamefont {Winchester}, \citenamefont {Gu},
  \citenamefont {Liu}, \citenamefont {Zhang}, \citenamefont {Wang},
  \citenamefont {Li}, \citenamefont {Chen}, \citenamefont {Eom}, \citenamefont
  {Schlom},\ and\ \citenamefont {Pan}}]{FE-Swicth}%
  \BibitemOpen
  \bibfield  {author} {\bibinfo {author} {\bibfnamefont {C.~T.}\ \bibnamefont
  {Nelson}}, \bibinfo {author} {\bibfnamefont {P.}~\bibnamefont {Gao}},
  \bibinfo {author} {\bibfnamefont {J.~R.}\ \bibnamefont {Jokisaari}}, \bibinfo
  {author} {\bibfnamefont {C.}~\bibnamefont {Heikes}}, \bibinfo {author}
  {\bibfnamefont {C.}~\bibnamefont {Adamo}}, \bibinfo {author} {\bibfnamefont
  {A.}~\bibnamefont {Melville}}, \bibinfo {author} {\bibfnamefont {S.-H.}\
  \bibnamefont {Baek}}, \bibinfo {author} {\bibfnamefont {C.~M.}\ \bibnamefont
  {Folkman}}, \bibinfo {author} {\bibfnamefont {B.}~\bibnamefont {Winchester}},
  \bibinfo {author} {\bibfnamefont {Y.}~\bibnamefont {Gu}}, \bibinfo {author}
  {\bibfnamefont {Y.}~\bibnamefont {Liu}}, \bibinfo {author} {\bibfnamefont
  {K.}~\bibnamefont {Zhang}}, \bibinfo {author} {\bibfnamefont
  {E.}~\bibnamefont {Wang}}, \bibinfo {author} {\bibfnamefont {J.}~\bibnamefont
  {Li}}, \bibinfo {author} {\bibfnamefont {L.-Q.}\ \bibnamefont {Chen}},
  \bibinfo {author} {\bibfnamefont {C.-B.}\ \bibnamefont {Eom}}, \bibinfo
  {author} {\bibfnamefont {D.~G.}\ \bibnamefont {Schlom}}, \ and\ \bibinfo
  {author} {\bibfnamefont {X.}~\bibnamefont {Pan}},\ }\href {\doibase
  10.1126/science.1206980} {\bibfield  {journal} {\bibinfo  {journal}
  {Science}\ }\textbf {\bibinfo {volume} {334}},\ \bibinfo {pages} {968}
  (\bibinfo {year} {2011})}\BibitemShut {NoStop}%
\bibitem [{\citenamefont {Cao}(2005)}]{FE-SwitchStraLim}%
  \BibitemOpen
  \bibfield  {author} {\bibinfo {author} {\bibfnamefont {W.}~\bibnamefont
  {Cao}},\ }\href {\doibase https://doi.org/10.1038/nmat1506} {\bibfield
  {journal} {\bibinfo  {journal} {Nature Materials}\ }\textbf {\bibinfo
  {volume} {4}},\ \bibinfo {pages} {727} (\bibinfo {year} {2005})}\BibitemShut
  {NoStop}%
\bibitem [{\citenamefont {Lotnyk}\ \emph {et~al.}(2016)\citenamefont {Lotnyk},
  \citenamefont {Ross}, \citenamefont {Bern{\"u}tz}, \citenamefont
  {Thelander},\ and\ \citenamefont {Rauschenbach}}]{STO-trivialPos}%
  \BibitemOpen
  \bibfield  {author} {\bibinfo {author} {\bibfnamefont {A.}~\bibnamefont
  {Lotnyk}}, \bibinfo {author} {\bibfnamefont {U.}~\bibnamefont {Ross}},
  \bibinfo {author} {\bibfnamefont {S.}~\bibnamefont {Bern{\"u}tz}}, \bibinfo
  {author} {\bibfnamefont {E.}~\bibnamefont {Thelander}}, \ and\ \bibinfo
  {author} {\bibfnamefont {B.}~\bibnamefont {Rauschenbach}},\ }\href {\doibase
  10.1038/srep26724} {\bibfield  {journal} {\bibinfo  {journal} {Scientific
  Reports}\ }\textbf {\bibinfo {volume} {6}},\ \bibinfo {pages} {26724 (9pp.)}
  (\bibinfo {year} {2016})}\BibitemShut {NoStop}%
\bibitem [{\citenamefont {Kramers}(1940)}]{Kramers}%
  \BibitemOpen
  \bibfield  {author} {\bibinfo {author} {\bibfnamefont {A.~H.}\ \bibnamefont
  {Kramers}},\ }\href@noop {} {\bibfield  {journal} {\bibinfo  {journal}
  {Pysica}\ }\textbf {\bibinfo {volume} {7}},\ \bibinfo {pages} {284} (\bibinfo
  {year} {1940})}\BibitemShut {NoStop}%
\bibitem [{\citenamefont {Mani}, \citenamefont {Chang},\ and\ \citenamefont
  {Ponomareva}(2013)}]{DynaStatFE}%
  \BibitemOpen
  \bibfield  {author} {\bibinfo {author} {\bibfnamefont {B.}~\bibnamefont
  {Mani}}, \bibinfo {author} {\bibfnamefont {C.}~\bibnamefont {Chang}}, \ and\
  \bibinfo {author} {\bibfnamefont {I.}~\bibnamefont {Ponomareva}},\
  }\href@noop {} {\bibfield  {journal} {\bibinfo  {journal} {Physical Review}\
  }\textbf {\bibinfo {volume} {B88}},\ \bibinfo {pages} {064306} (\bibinfo
  {year} {2013})}\BibitemShut {NoStop}%
\end{thebibliography}
%

\end{document}